\documentclass[%
aps,
pre,
superscriptaddress,
showpacs,showkeys,
a4paper,
12pt,
longbibliography,
reprint,
notitlepage
]{revtex4-1}
\usepackage[english]{babel}
\usepackage{amssymb,amsmath,stmaryrd,array}

\usepackage{graphicx}
\usepackage{subfig}

\makeatletter

\newcommand{\avr}[1]{\ensuremath{\langle{#1}\rangle}}

\newcommand{\cnj}[1]{{#1}^{\ast}}

\renewcommand{\Re}{\mathop{\rm Re}\nolimits}
\renewcommand{\Im}{\mathop{\rm Im}\nolimits}

\newcommand{\mum}{$\mu$m}
\newcommand{\dega}{\ensuremath{^\circ}}

 \newcommand{\bs}[1]{\boldsymbol{#1}}
 \newcommand{\vc}[1]{\mathbf{#1}}
 
 \newcommand{\uvc}[1]{\hat{\mathbf{#1}}}
 
 \newcommand{\ind}[1]{\mathrm{#1}}

\newcommand{\dd}{\mathrm{d}}

 \newcommand{\e}{\mathrm{e}}

\newcommand{\eff}{\mathrm{eff}}

\makeatother

\begin{document}
\DeclareGraphicsExtensions{.pdf,.eps,.png}

\title{Modulation of unpolarized light in planar aligned
subwavelength-pitch deformed-helix ferroelectric liquid crystals} 

\author{Vladimir~V.~Kesaev}
 \email[Email address: ]{vladimir.kesaev@gmail.com}
\affiliation{%
Lebedev Physical Institute,
Leninsky Prospekt 53, 119991 Moscow, Russia
 }

\author{Alexei~D.~Kiselev}
\email[Email address: ]{alexei.d.kiselev@gmail.com}
\affiliation{%
 Saint Petersburg National Research University of Information Technologies,
 Mechanics and Optics (ITMO University),
 Kronverskyi Prospekt 49,
 197101 Saint Petersburg, Russia}

\author{Evgeny~P.~Pozhidaev}
 \email[Email address: ]{epozhidaev@mail.ru}
\affiliation{%
Lebedev Physical Institute,
Leninsky Prospekt 53, 119991 Moscow, Russia
 }

\date{\today}

\begin{abstract}
  We 
study the electro-optic properties of subwavelength-pitch
  deformed-helix ferroelectric liquid crystals (DHFLC) 
illuminated with unpolarized light.
In the experimental setup based on the Mach-Zehnder
interferometer, it was observed
that the reference and the sample beams being
 both unpolarized produce the interference pattern which
  is insensitive to rotation of in-plane optical axes of the DHFLC
  cell.  
We find that the
  field induced shift of the interference fringes 
can be described in terms of 
the electrically dependent
  Pancharatnam relative phase determined by 
the averaged phase shift, 
whereas the visibility of the fringes is solely dictated by the
  phase retardation.
\end{abstract}

 \maketitle

\section{Introduction}
\label{sec:intro}

Liquid crystal (LC) spatial light modulators (SLMs) are
known to be the key elements 
widely employed
to modulate amplitude,
phase, or polarization of light waves
in space and time~\cite{Efron:bk:1995}. 
Ferroelectric liquid crystals
(FLCs) represent most promising chiral liquid crystal material 
which is characterized by very fast response time
(a detailed description of FLCs can be found, e.g.,
in~\cite{Oswald:bk:2006}).
However, most of the FLC modes are not suitable for 
phase-only
modulation devices because their optical axis sweeps in 
the plane of the cell substrate
producing undesirable changes in the polarization state of the incident light. 

In order to get around
the optical axis switching problem
the system consisting of a FLC half-wave plate sandwiched between two 
quarter-wave plates was suggested in~\cite{Love:optcom:1994}.
But, in this system, the $2\pi$ phase modulation requires the smectic
tilt angle to be equal to 45$^\circ$. 
This value is a real challenge for the material science
and, in addition,
the response time dramatically increases when the tilt angle
grows up to 45$^\circ$~\cite{Barnik:mclc:1987,Pozhidaev:jetp:1988}. 

An alternative approach proposed in~\cite{Kiselev:pre:2013,Kiselev:ol:2014} 
is based on the orientational Kerr effect in a vertically aligned deformed helix
ferroelectric LC (DHFLC) with subwavelength helix pitch. 
This effect is characterized by fast 
and, under certain conditions~\cite{Blinov:pre:2005,Pozhidaev:jsid:2012},
hysteresis-free electro-optics that preserves ellipticity of the incident light.

An important point is that
all the above mentioned studies
deal with the case of linearly polarized
illumination where
the incident beam is fully polarized.
In this paper,
we will go beyond this limitation
and examine 
the electro-optic behavior of 
the planar aligned DHFLC cells illuminated with
unpolarized light
as the way to obtain
the response insensitive
to the effect of 
electric-field-induced rotation
of in-plane optical axes
which was found to 
be responsible for the presence of amplitude 
modulation~\cite{Kiselev:pre:2015}. 
The paper is organized as follows.

In Sec.~\ref{sec:experiment},
after introducing
the geometry of the DHFLC cell
and discussing the orientational Kerr effect,
we describe the samples and
our experimental setup which is based
on the Mach-Zehnder interferometer.
The experimental results
are interpreted theoretically
in Sec.~\ref{sec:theory}.
Finally, in Sec.~\ref{sec:conclusion}
we draw the results together and make some concluding
remarks.
Details on the effective dielectric tensor
of DHFLC cells
are given in the Appendix.

\section{Experiment}
\label{sec:experiment}

\begin{figure}[!tbh]
\centering
\resizebox{85mm}{!}{\includegraphics*{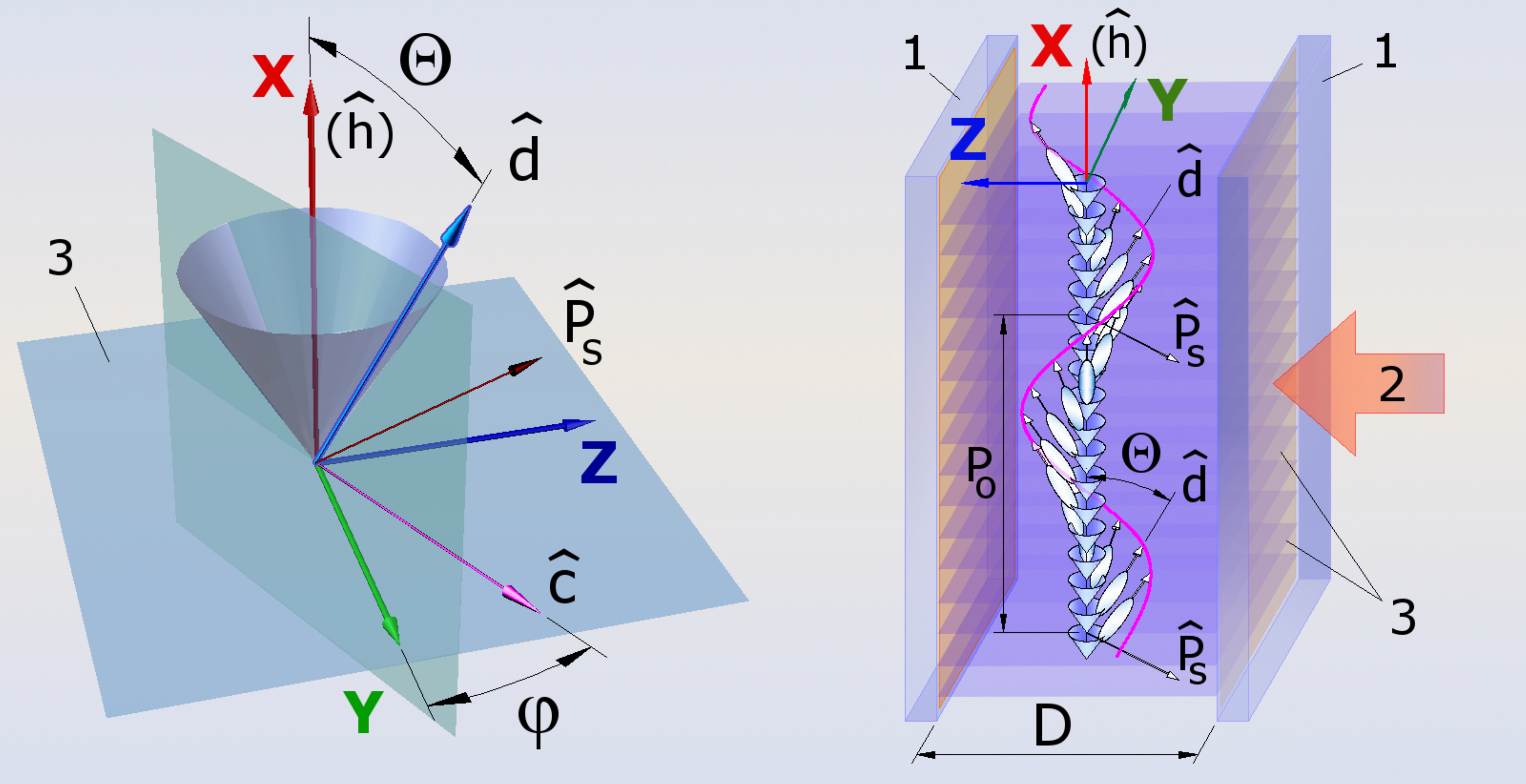}}
\caption{%
Geometry of 
smectic cone
(left)
and DHFLC cell
(right).
1~--~cell substrates;
2~--~incident light beam;
3~--~smectic layers.
}
\label{fig:cell}
\end{figure}

We consider a DHFLC layer of thickness $D$
with the $z$ axis 
normal to the bounding surfaces:
$z=0$ and $z=D$ (see Fig.~\ref{fig:cell}).
The geometry of
a \textit{uniform lying} DHFLC helix
with subwavelength pitch
where the helix (twisting) axis
$\uvc{h}$
is directed along the $x$ axis
will be our primary concern. 
In this geometry depicted in Fig.~\ref{fig:cell}
the equilibrium orientational structure 
can be described as a helical twisting pattern
where FLC molecules align on average along
a local unit director
\begin{align}
&
\uvc{d}=
\cos\theta\,\uvc{h}+
\sin\theta\,\uvc{c},
\label{eq:director}
  \end{align}
where $\theta$ is the \textit{smectic tilt angle}; 
$\uvc{h}=\uvc{x}$ is the twisting axis normal to the smectic layers and
$\uvc{c}\perp\uvc{h}$ is the $c$-director.
The FLC director~\eqref{eq:director}
lies on the smectic cone 
depicted in Fig.~\ref{fig:cell}(left)
and rotates
in a helical fashion about a uniform twisting axis
$\uvc{h}$ forming the FLC helix. 
This rotation is described by
the azimuthal angle
around the cone $\phi$
that specifies
orientation of the $c$-director in the plane perpendicular to
$\uvc{h}$.  

Figure~\ref{fig:cell}(right)
illustrates
the uniform lying FLC helix
in the slab geometry 
with the smectic layers normal to the substrates
and
\begin{align}
&
\uvc{h}=\uvc{x},
\quad
\uvc{c}=\cos\phi\,\uvc{y}+
\sin\phi\,\uvc{z},
\quad
\vc{E}=E\,\uvc{z},
\label{eq:director-planar}    
  \end{align}
where $\vc{E}$ is the applied electric field
which is linearly coupled to
the \textit{spontaneous ferroelectric polarization}
\begin{align}
&
\vc{P}_s=P_s\uvc{p},
\quad
\uvc{p}=
\uvc{h}\times\uvc{c}=\cos\phi\,\uvc{z}-
\sin\phi\,\uvc{y},
\label{eq:pol-vector}
  \end{align}
where $\uvc{p}$ is
the \textit{polarization unit vector}.

For a biaxial FLC, the components of the dielectric tensor,
$\bs{\varepsilon}$,
are given by
\begin{align}
&
\epsilon_{i j}=
  \epsilon_{\perp}
\delta_{i j}+
(\epsilon_{1}-\epsilon_{\perp})\,
d_i d_j
+
(\epsilon_{2}-\epsilon_{\perp})\,
p_i p_j
\notag
\\
&
=
  \epsilon_{\perp}
[
\delta_{i j}+(r_1-1) d_i d_j
+(r_2-1) p_i p_j
],
\label{eq:diel-tensor}
\end{align}
where 
$i,j\in\{x,y,z\}$,
$\delta_{ij}$ is the Kronecker delta;
$d_i$ ($p_i$) is the $i$th component
of the FLC director [unit polarization vector]
given by Eq.~\eqref{eq:director} [Eq.~\eqref{eq:pol-vector}];
$r_1=\epsilon_1/\epsilon_\perp$ ($r_2=\epsilon_2/\epsilon_\perp$)
is the \textit{anisotropy (biaxiality) ratio}.
In the case of uniaxial anisotropy with $r_2=1$,
we have the two principal values of the dielectric tensor,
$\epsilon_2=\epsilon_{\perp}$ and
$\epsilon_{1}=\epsilon_{\parallel}$,
giving the ordinary (extraordinary) refractive index
 $n_{\perp}=\sqrt{\mu\epsilon_{\perp}}$
($n_{\parallel}=\sqrt{\mu\epsilon_{\parallel}}$),
where
the magnetic tensor of FLC 
is assumed to be isotropic with the magnetic permittivity $\mu$.

According to
Refs.~\cite{Kiselev:pre:2011,Kiselev:pre:2:2014,Kiselev:pre:2015},
optical properties of such cells
can be described by the effective dielectric tensor
of a homogenized DHFLC helical structure
(the results used in this paper 
are summarized in the Appendix).
As is shown in Fig.~\ref{fig:ellipsoids},
the zero-field 
($\mathbf{E}=0$)
dielectric tensor is
uniaxially anisotropic with 
the optical axis directed along the
twisting axis $\hat{\mathbf{h}}=\hat{\mathbf{x}}$.
The zero-field effective refractive indices of
extraordinary (ordinary) waves, $n_h$ ($n_p$),
generally depend
on the smectic tilt angle
$\theta$ and
the optical (high frequency) dielectric
constants
characterizing the FLC
material that enter the tensor
given in Eq.~\eqref{eq:diel-tensor}
(the expressions for $\epsilon_h=n_h^2$
and $\epsilon_p=n_p^2$
are given by
Eq.~\eqref{eq:epsilon_ph}
in the Appendix).

Referring to Fig.~\ref{fig:ellipsoids}, 
the electric-field-induced
anisotropy is generally biaxial
so that the dielectric tensor
is characterized by the three generally different principal values
(eigenvalues):
$\epsilon_{\pm}=n_{\pm}^2$ and $\epsilon_z=n_z^2$
(see Eqs.~\eqref{eq:epsilon_z} and~\eqref{eq:epsilon_pm}
in the Appendix).
The in-plane principal optical axes 
(eigenvectors)
\begin{align}
  \label{eq:d-pm}
  \hat{\mathbf{d}}_{+}=\cos\psi_d\,\hat{\mathbf{x}}+
\sin\psi_d\,\hat{\mathbf{y}},
\quad
 \hat{\mathbf{d}}_{-}=\hat{\mathbf{z}}\times  \hat{\mathbf{d}}_{+}
\end{align}
are rotated about the
vector of electric field, $\mathbf{E}\parallel \hat{\mathbf{z}}$,
by the azimuthal angle $\psi_{\mathrm d}$
(see Eq.~\eqref{eq:psi_d} in the Appendix).
 In the low electric field  region,
the electric field dependence of the
angle $\psi_{\mathrm d}$ is
approximately linear:
$\psi_{\mathrm d}\propto E$,
whereas the electrically induced
part of the principal refractive indices,
$n_{\pm}$ and $n_z$,
is typically dominated by the
Kerr-like nonlinear terms proportional to $E^2$
(see, e.g., equations (10)--(18) in
Ref.~\cite{Kiselev:pre:2015}).
This effect ~---~
the so-called \textit{orientational Kerr effect}~---~
is caused by
the electrically induced distortions of the helical
structure.
It is governed by the effective dielectric tensor
of a nanostructured chiral smectic liquid crystal
defined through averaging over 
the FLC orientational 
structure~\cite{Kiselev:pre:2013,Kiselev:ol:2014,Kiselev:pre:2011,Kiselev:pre:2:2014}.

\begin{figure}[!tbh]
\centering
\resizebox{70mm}{!}{\includegraphics*{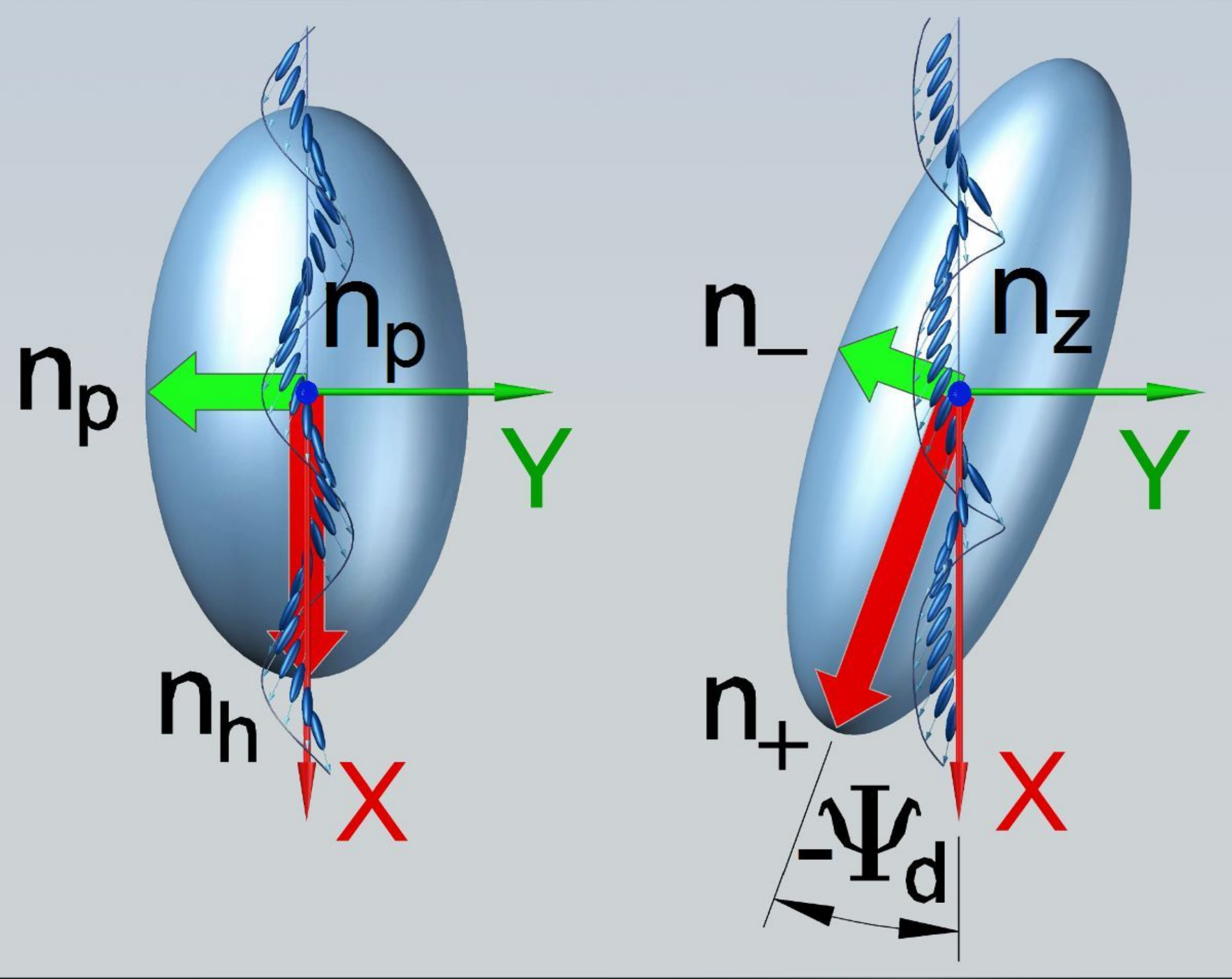}}
\caption{%
Ellipsoids of effective refractive indices of a short-pitch
DHFLC cell. 
Left: at $\mathbf{E}=0$, 
the field-free effective ellipsoid is
uniaxially anisotropic with the optical axis parallel to the helix
axis (the $x$ axis).
The refractive indices $n_h$ and $n_p$ 
correspond to the optical axes along and perpendicular to 
the helix axis, respectively. 
Right: applying an electric field across the cell, makes the
optical anisotropy biaxial with the two in-plane 
optical axes rotated by the angle $\psi_d\propto E$ 
about the electric field vector $\mathbf{E}$.
}
\label{fig:ellipsoids}
\end{figure}

In our experiments we have used the FLC mixture FLC-624
(from P.N. Lebedev Physical Institute of Russian Academy of
Sciences) as the material for the DHFLC layer. 
The FLC-624 is a
mixture of the FLC-618 
(the chemical structure 
can be found in~\cite{Kiselev:ol:2014}) 
and the well-known nematic liquid crystal 5CB. 
The weight concentrations in the mixture are
98\% of FLC-618 and 2\% of 5CB. 
The phase
transition sequence of the FLC-624 during heating from the solid
crystalline phase is 
$
\mathrm{Cr}\xrightarrow{+20^{\circ}\mathrm{C}}\mathrm{Sm}C^{\star}\xrightarrow{+60^{\circ}\mathrm{C}}\mathrm{Sm}A^{\star}\xrightarrow{+116^{\circ}\mathrm{C}} \mathrm{Iso},
$
whereas at cooling down from isotropic phase crystallization
occurs around +6$^\circ$C. 
At room temperature ($T=23^{\circ}$C),
the spontaneous polarization
$P_s$
and the FLC helix pitch $p_0$
are
$185$~nC/cm$^2$
and
$210$~nm, respectively.

The FLC-624 layer is sandwiched between two glass substrates
covered by indium tin oxide (ITO) and aligning films with 
the thickness
$20$~nm and the gap is fixed by spacers at 
$D\approx 53$~$\mu$m. 
The FLC alignment technique 
that provides high-quality chevron-free
planar alignment with smectic layers perpendicular to 
the substrates is detailed in Ref.~\cite{Kiselev:pre:2015}.

\begin{figure}[!tbh]
\centering
\resizebox{84mm}{!}{\includegraphics*{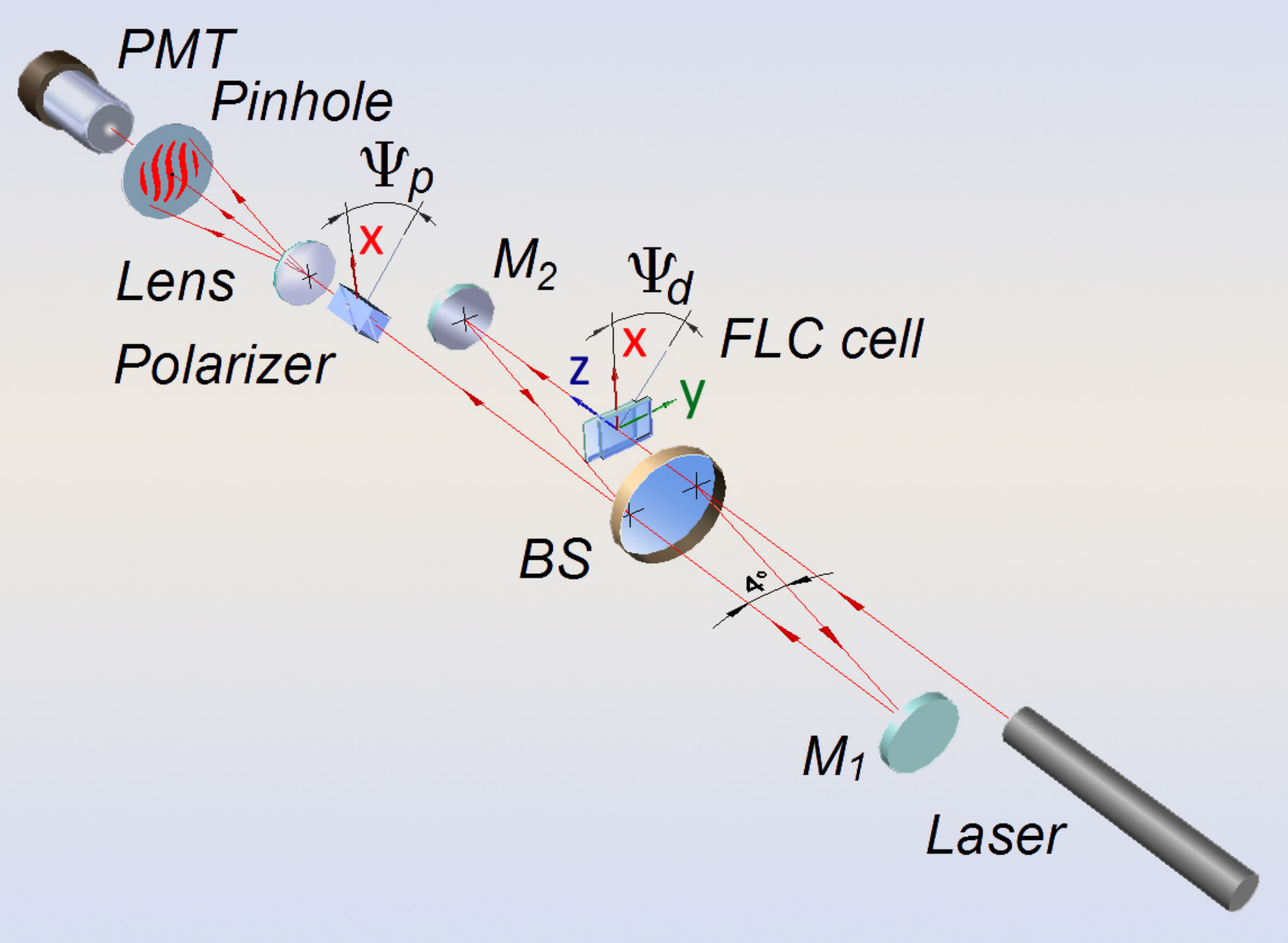}}
\caption{%
Experimental setup 
based on a Mach-Zehnder two-arm interferometer:
BS is the beam splitter;
M$_1$ and M$_2$ are the mirrors;
PMT is the photomultiplier tube.
}
\label{fig:mz}
\end{figure}

 Electro-optical studies
of DHFLC cells
are typically carried out by
measuring  
the transmittance of normally incident 
light passing through crossed polarizers.
By contrast, our experimental setup
shown in Fig.~\ref{fig:mz}
is based on a Mach-Zehnder two-beam interferometer
where the FLC cell is placed in the path of the sample beam.

A helium-neon laser with the wavelength of $632.8$~nm 
was used as a source of unpolarized light. 
A beam splitter (BS) divides a collimated unpolarized laser light
into two beams, the reference and the sample beams,
which, after reflection
at the mirrors $M_1$ and $M_2$, 
are recombined at the semireflecting surface
of  the beam splitter (BS). 
The interfering beams
emerging from the interferometer 
optionally pass through an output polarizer 
with the transmission axis azimuth $\psi_p$
and then are projected by the lens 
on to a screen with a pinhole. 
After passing the pinhole,
light  is collected by a photomultiplier (PMT)
used in the linear regime as a photodetector. 

\begin{figure}[!tbh]
\centering
\resizebox{70mm}{!}{\includegraphics*{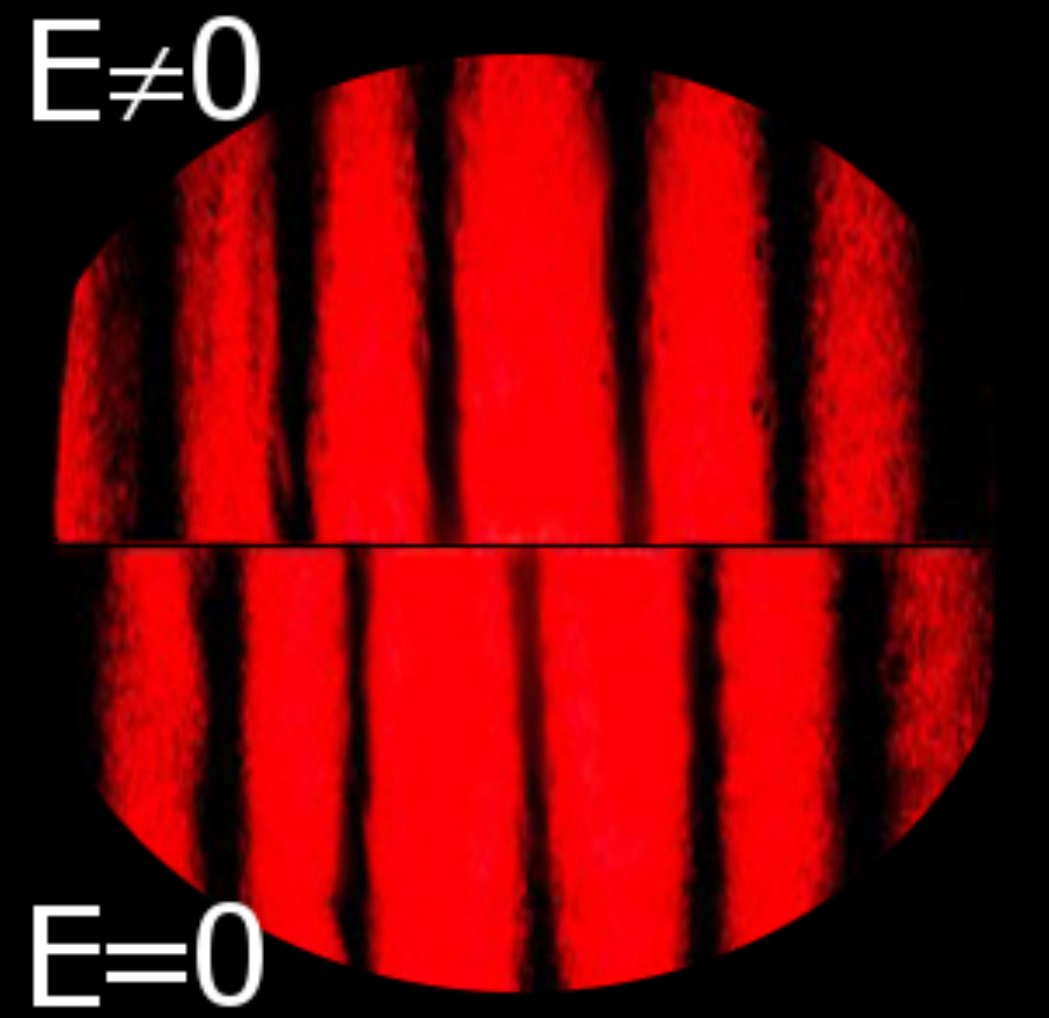}}
\caption{%
The interference fringes
at $E\ne 0$ (top) and $E=0$
(bottom).
}
\label{fig:fringes}
\end{figure}

\begin{figure}[!tbh]
\centering
\resizebox{70mm}{!}{\includegraphics*{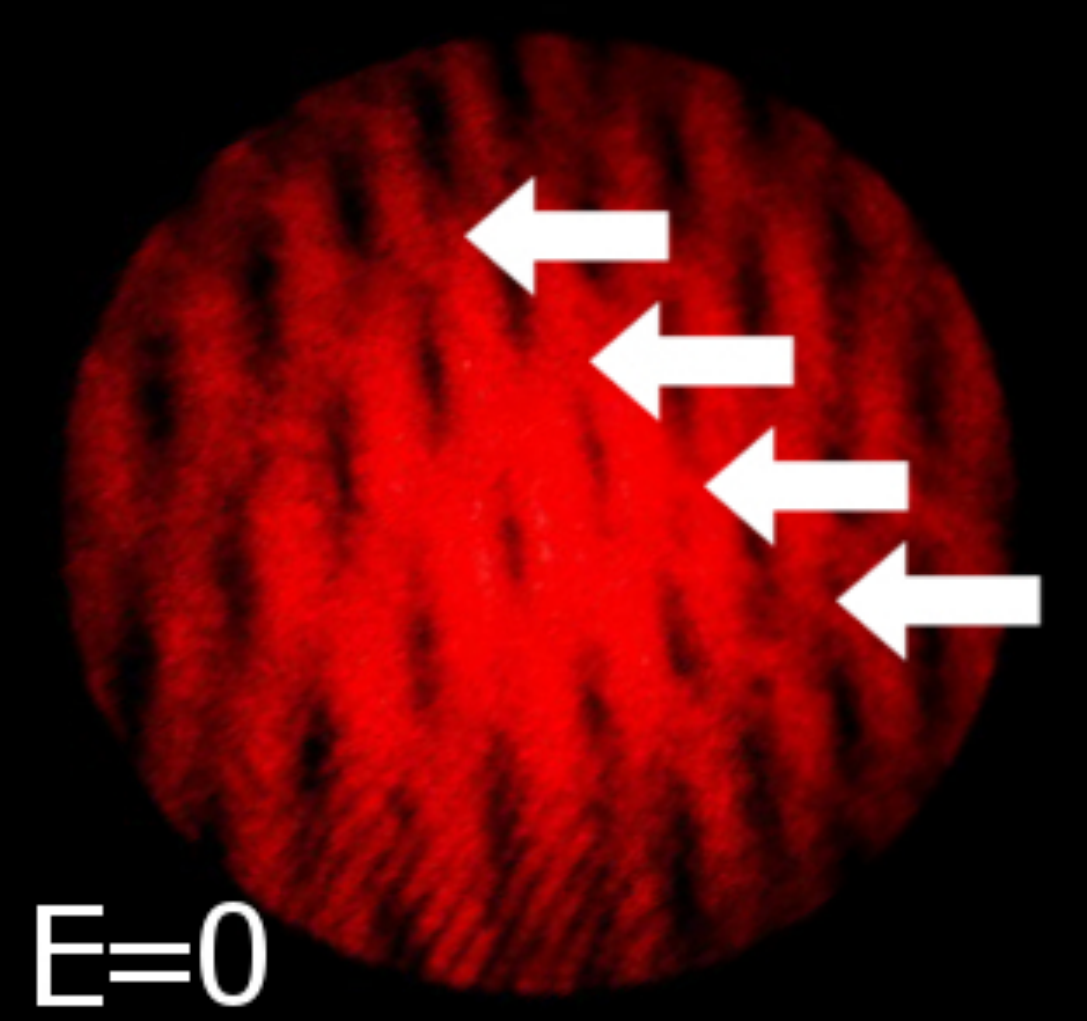}}
\caption{%
The interferogram obtained without
the output polarizer
shows the half-wave fringes indicated by arrows.
}
\label{fig:hw-fringes}
\end{figure}

The interferometer was adjusted 
to obtain the fringes of equal thickness
shown in Figs.~\ref{fig:fringes} and~\ref{fig:hw-fringes}. 
The period of the interference pattern
was 120 times larger than the pinhole diameter
and thus our
measurements of the light intensity 
 were performed with
an accuracy less than 0.7\%. 
In order to minimize 
the polarizing effects of Fresnel reflections, 
all the directions of incidence were close to
the normal (deviations from the normal were less than $2^\circ$)
so that the measured values of the degree of polarization
of both beams were below $10^{-4}$.

In order to facilitate 
data processing for signals registered by PMT,
the mirrors $M_1$ and $M_2$ were fine tuned
so as to bring the pinhole into coincidence
with the position of an intensity minimum
of the field-free interference pattern. 
The experimental results
measured for triangular wave-form of driving voltage
with the frequency $f=50$~Hz
(the voltage applied across the DHFLC cell ranges from
$-40$~V to $+40$~V) 
are
presented in Figs.~\ref{fig:inten-total} and~\ref{fig:inten-polarizer}.
These figures clearly indicate that 
the applied voltage results in 
modulation of 
the light intensity
no matter whether the output wavefield 
passes through the polarizer or not.
Another important effect is
the electric field induced
shift of the interference fringes
which is
illustrated in
Fig.~\ref{fig:fringes}.

\section{Theory}
\label{sec:theory}

Theoretically,
the above findings can be interpreted
in terms of the output beam written 
as the sum of the vector amplitudes
\begin{align}
  \label{eq:E_total} 
\mathbf{E}=\mathbf{E}_s+\mathbf{E}_r
\end{align}
representing
the
light transmitted through the FLC cell
(the sample beam),
$\mathbf{E}_s=\begin{pmatrix}
  E_x^{(s)}\\
E_y^{(s)}
\end{pmatrix}$,
and the reference beam,
$\mathbf{E}_r=\begin{pmatrix}
  E_x^{(r)}\\
E_y^{(r)}
\end{pmatrix}
=
\mathbf{T}_r\mathbf{E}_0=\mathrm{e}^{i\Phi_{0}}\mathbf{E}_0$,
where $\mathbf{T}_r=\mathrm{e}^{i\Phi_{0}}\mathbf{I}_2$
and $\mathbf{I}_2$
is the $2\times 2$ unity matrix.

 The vector amplitudes of incident and transmitted waves,
$\mathbf{E}_0$ and $\mathbf{E}_{s}$,
are related through the standard input-output relation
\begin{align}
  \label{eq:input-output}
  \mathbf{E}_{s}=\mathbf{T}_s\mathbf{E}_0,
\end{align}
where $\mathbf{T}_s$ is the transmission matrix
that,
for the case of normal incidence,
can be easily obtained
from the general results of Refs.~\cite{Kiselev:pra:2008,Kiselev:pre:2:2014}
in the form:
\begin{align}
&
  \label{eq:T-norm}
 \mathbf{T}_s
=
t_{+}\uvc{d}_{+}\otimes\cnj{\uvc{d}}_{+}+
t_{-}\uvc{d}_{-}\otimes\cnj{\uvc{d}}_{-},
\\  
 \label{eq:t-pm}
&
  t_{\pm}=|t_{\pm}|\e^{i\Phi_{\pm}}=\frac{1-\rho_{\pm}^2}{%
1-\rho_{\pm}^2\exp(2in_{\pm}h)
}\exp(i n_{\pm} h),
\end{align}
where
$\rho_{\pm}=(n_{\pm}-n_{\ind{m}})/(n_{\pm}+n_{\ind{m}})$ 
is the Fresnel reflection coefficient;
$n_{\ind{m}}$ is the refractive index of the ambient medium;
$h=k_{\mathrm{vac}} D$ is the thickness parameter;
$k_{\mathrm{vac}}=\omega/c$ is the free-space wavenumber;
and an asterisk will indicate complex conjugation.
When the reflection coefficients, $\rho_{\pm}$, are small, 
the transmission coefficients,
$t_{+}$ and $t_{-}$,
can be approximated by the well-known formulas 
\begin{align}
  \label{eq:t_pm_approx}
&
  t_{\pm}\approx \exp(i\Phi_{\pm}), 
\quad
\Phi_{\pm}=n_{\pm} h.
\end{align}
and, in this reflectionless approximation, 
the transmission matrix is unitary.

The beam emerging from the
interferometer
and the incident light are
characterized by the $2\times 2$
equal-time coherence
matrices~\cite{Mandl:bk:1995,Brosseau:bk:1998},
$\mathbf{M}$ and $\mathbf{M}_0$,
with the elements
$\mathbf{M}_{\alpha\beta}=
\bigl\langle
E_{\alpha} E_{\beta}^{\ast}
\bigr\rangle$
and 
$\mathbf{M}_{\alpha\beta}^{(0)}=
\bigl\langle
E_{\alpha}^{(0)}
\bigl[E_{\beta}^{(0)}\bigr]^{\ast}
\bigr\rangle$,
respectively.
From the general relation linking these matrices
$\mathbf{M}=
\mathbf{T}\mathbf{M}_0 \mathbf{T}^{\dagger}$,
where 
$\mathbf{T}=\mathbf{T}_s+\mathbf{T}_r$
and
a dagger will denote
Hermitian conjugation,
we have the coherence matrix of the output
wavefield~(\ref{eq:E_total}) given by
\begin{align}
  \label{eq:MvsM0}
\mathbf{M}=
\mathbf{M}_r+\mathbf{M}_s+
\mathbf{M}_i,
\quad
\mathbf{M}_i=
\mathbf{T}_s\mathbf{M}_0\mathbf{T}_r^{\dagger}
+\mathbf{T}_r\mathbf{M}_0\mathbf{T}_s^{\dagger},
\end{align}
where 
$\mathbf{M}_{r}=
\mathbf{T}_{r}\mathbf{M}_0\mathbf{T}_{r}^{\dagger}=\mathbf{M}_0$
is the coherence matrix of the reference beam, 
$\mathbf{M}_{s}=
\mathbf{T}_{s}\mathbf{M}_0\mathbf{T}_{s}^{\dagger}$
is the coherence matrix of 
the light transmitted through the DHFLC cell
and $\mathbf{M}_i$ is the interference term.

For the case where 
the input light is unpolarized
and
its coherency matrix is
proportional to the unity matrix,
$\mathbf{M}_{0}=\dfrac{I_0}{2} \mathbf{I}_2\equiv \mathbf{M}_u$,
the coherence matrix~(\ref{eq:MvsM0}) 
takes the form:
\begin{align}
&
  \label{eq:M}
\frac{1}{I_0} \mathbf{M}=\frac{m_0}{2}\mathbf{I}_2
+\frac{m_p}{2}
\begin{pmatrix}
  \cos 2\psi_\mathrm{d}&\sin 2\psi_\mathrm{d}\\
\sin 2\psi_\mathrm{d}& -\cos 2\psi_\mathrm{d}
\end{pmatrix},
\\
&
  \label{eq:intensity}
I/I_0 \equiv m_0=
1+\frac{|t_{+}|^2+|t_{-}|^2}{2}+\mathrm{Re}[\mathrm{e}^{-i\Phi_{0}}(t_{+}+t_{-})]
\notag
\\
&
\approx
2\{1+\cos(\Phi-\Phi_0)\cos(\Delta\Phi)\},
\\
&
  \label{eq:Phi}
\Phi=(\Phi_{+}+\Phi_{-})/2,
\quad
\Delta\Phi=(\Phi_{+}-\Phi_{-})/2,
\\
&
\label{eq:s_p}
m_p=\frac{|t_{+}|^2-|t_{-}|^2}{2}+
\mathrm{Re}[\mathrm{e}^{-i\Phi_{0}}(t_{+}-t_{-})]
\notag
\\
&
\approx -2 \sin(\Phi-\Phi_0)\sin(\Delta\Phi),
\end{align}
where
$I=\langle\mathbf{E}\cdot\mathbf{E}^{\ast}\rangle$
is the total intensity
of the beams exiting the interferometer;
$\Phi$ is the \textit{averaged phase shift};
$2\Delta\Phi=(n_{+}-n_{-})h$
is the difference in optical path of the ordinary and extraordinary
waves known as the \textit{phase
  retardation}.

The approximate expressions for
the \textit{intensity and polarization
parameters},
$m_0$ and $m_p$, are derived
using 
the approximation given by
Eq.~\eqref{eq:t_pm_approx}.
This is the case where
the matrix $\mathbf{T}_s$ is unitary
and, similar to the reference beam,
the light field emerging from the FLC cell
is unpolarized:
$\mathbf{M}_s=\mathbf{M}_r=\mathbf{M}_u$.
So, the only electric field dependent
contribution to
the coherence matrix of the total
wavefield $\mathbf{M}$
(see Eq.~\eqref{eq:MvsM0})
comes from
the interference term $\mathbf{M}_i$.

This term is responsible for 
the following two effects: 
(a)~electrically induced
modulation of the intensity
described by Eq.~\eqref{eq:intensity};
and (b)~the total wavefield~\eqref{eq:E_total}
is partially polarized with 
the degree of polarization $P=|m_p|/m_0$
and
the Stokes parameters 
\begin{align}
  \label{eq:stokes-output}
(S_1,S_2,S_3)=
I_0
m_p(\cos 2\psi_{\mathrm{d}},\sin
2\psi_{\mathrm{d}},0)\equiv I_0 m_p\uvc{s}_{d}  
\end{align}
varying with applied electric field. 
The latter, in particular, implies that
the intensity of
light that after the interferometer passes through a linear
polarizer  
\begin{align}
  \label{eq:inten-polarizer}
I_p(\psi_p)/I_0=
[m_0+m_p \cos 2(\psi_{\mathrm{d}}-\psi_p)]/2
\end{align}
depends on orientation 
of the polarizer transmission axis
specified by
the azimuthal angle 
$\psi_p$.

From~\eqref{eq:intensity}, 
electric-field-induced modulation of
the intensity 
does not depend on
orientation
of the optical axes 
(the azimuthal angle $\psi_{\mathrm{d}}$)
and
will occur even if the cell is optically
isotropic and $\Phi_{+}=\Phi_{-}=\Phi$.
Interestingly, 
this effect can be interpreted
in terms of the \textit{Pancharatnam phase},
$\Phi_P$.

This phase 
has a long history
dating back to the original paper
by Pancharatnam~\cite{Pancharatnam:proc:1956}
(see also a collection of important papers~\cite{Shapere:bk:1989})
and can naturally be defined as 
the phase acquired  by a light wave
as it evolves along a path in
the space of polarization states.
In our case,
the Pancharatnam phase is 
generated by evolution of
mixed polarization states
governed by the
transmission matrix $\mathbf{T}_s(h)$
(a recent discussion of the Pancharatnam phase
for pure and mixed states 
in optics can be found, e.g., 
in~\cite{Hariharan:progr_opt:2005,
Barberena:pra:2016}).

To be more specific, let us consider 
a  partially polarized input beam
with the degree of polarization equal to
$P_0$
and
the normalized coherency matrix
$\rho_0=\mathbf{M}_0/I_0$
($\mathrm{Tr}\rho_0=1$)
which is known to play the role of
the density matrix describing
the mixed polarization state.
This matrix can generally be expressed
in terms of the eigenstates
as follows
\begin{align}
  \label{eq:rho_0}
  \rho_0=
\frac{1+P_0}{2}
\uvc{f}_{+}\otimes\cnj{\uvc{f}}_{+}
+\frac{1-P_0}{2}
\uvc{f}_{-}\otimes\cnj{\uvc{f}}_{-},
\end{align}
where the eigenpolarization vectors,
$\uvc{f}_{+}$ and $\uvc{f}_{-}$,
form the orthonormal basis
that meets the orthogonality conditions:
$\cnj{\uvc{f}}_{\mu}\cdot\uvc{f}_{\nu}=\delta_{\mu\nu}$
($\delta_{\mu\nu}$ is the Kronecker delta).
Similar to Eq.~\eqref{eq:stokes-output},
the Stokes vector of the mixed state~\eqref{eq:rho_0}
can be written in the form:
\begin{align}
  \label{eq:stokes-input}
  (S_1^{(0)},S_2^{(0)},S_3^{(0)})=
I_0
P_0 \uvc{s}_{0},
\end{align}
 where $\uvc{s}_{0}$ is the normalized unit Stokes
vector characterizing the partially polarized input wave.

In accordance with
the interferometry based 
approach~\cite{Sjoqvist:prl:2000,Tong:prl:2004}, 
this is the interference part of the intensity, $I_i=\mathrm{Tr}\mathbf{M}_i$,
determined by the interference term of the coherence 
matrix~\eqref{eq:MvsM0}
\begin{align}
  \label{eq:interference-intensity}
  I_i/I_0= 2\Re [\e^{-i\Phi_0} F_P]=2V\cos(\Phi_P-\Phi_0)
\end{align}
that gives the \textit{Pancharatnam phase},
$\Phi_P$, and
the \textit{visibility} of the interference pattern, 
$V$. These are thus defined by 
the averaged transmission matrix through
the general relation
\begin{align}
  \label{eq:PhiP}
F_P=\avr{\mathrm{T}_s(h)}_0\equiv\mathrm{Tr}[\mathbf{T}_s(h)\rho_0]=
V\exp(i\Phi_P).
\end{align}
For the mixed state~\eqref{eq:rho_0}
and the transmission matrix~\eqref{eq:T-norm},
it is not difficult to derive
the expression for $F_P$ 
which can be conveniently written in the
following form:
\begin{align}
  \label{eq:F_P}
&
  F_P=[\tau_{+}+\tau_{-} P_0\cos\psi_0],
\:
\cos\psi_0=\uvc{s}_0\cdot\uvc{s}_d,
\\
&
\label{eq:alpha_pm}
\tau_{\pm}=(t_{+}\pm t_{-})/2.
\end{align}

In the case of unpolarized light  
and unitary transmission matrix [see Eq.~\eqref{eq:t_pm_approx}]
where  $P_0=0$ and $\tau_{+}\approx \cos(\Delta\Phi)\exp(i\Phi)$,
the formulas for  the Pancharatnam phase and the visibility
assume the simplified form:
\begin{align}
  \label{eq:PhiP-unp}
&
\Phi_P=\arg F_P\approx
\begin{cases}
  \Phi, & \cos(\Delta\Phi)>0\\
  \Phi+\pi, & \cos(\Delta\Phi)<0\\
\end{cases}
\\
&
  \label{eq:V-unp}
V=|F_P|\approx |\cos(\Delta\Phi)|.
\end{align}
It should be stressed that
the phase $\Phi_P$ defined by 
the relations~\eqref{eq:PhiP}--\eqref{eq:alpha_pm}
is the total relative phase
between the beams before and after the cell.

For unpolarized light with $P_0=0$, 
the Pancharatnam phase~(\ref{eq:PhiP-unp}) 
is governed by the average of
in-plane refractive indices: $(n_{+}+n_{-})/2$, 
whereas
the visibility~(\ref{eq:V-unp})  
is determined by
the retardation phase
$2\Delta\Phi$.
The intensity of the
interference pattern~(\ref{eq:intensity})
can now be recast into the simple form  
\begin{align}
  I=\mathrm{Tr}\,\mathbf{M}=2I_0[1+V\cos(\Phi_P-\Phi_0)]
\end{align}
which is
manifestly independent of optical axes orientation and shows that the
electrically induced shift of the interferogram is solely dictated by
the Pancharatnam phase.

\begin{figure}[!tbh]
\centering
\resizebox{80mm}{!}{\includegraphics*{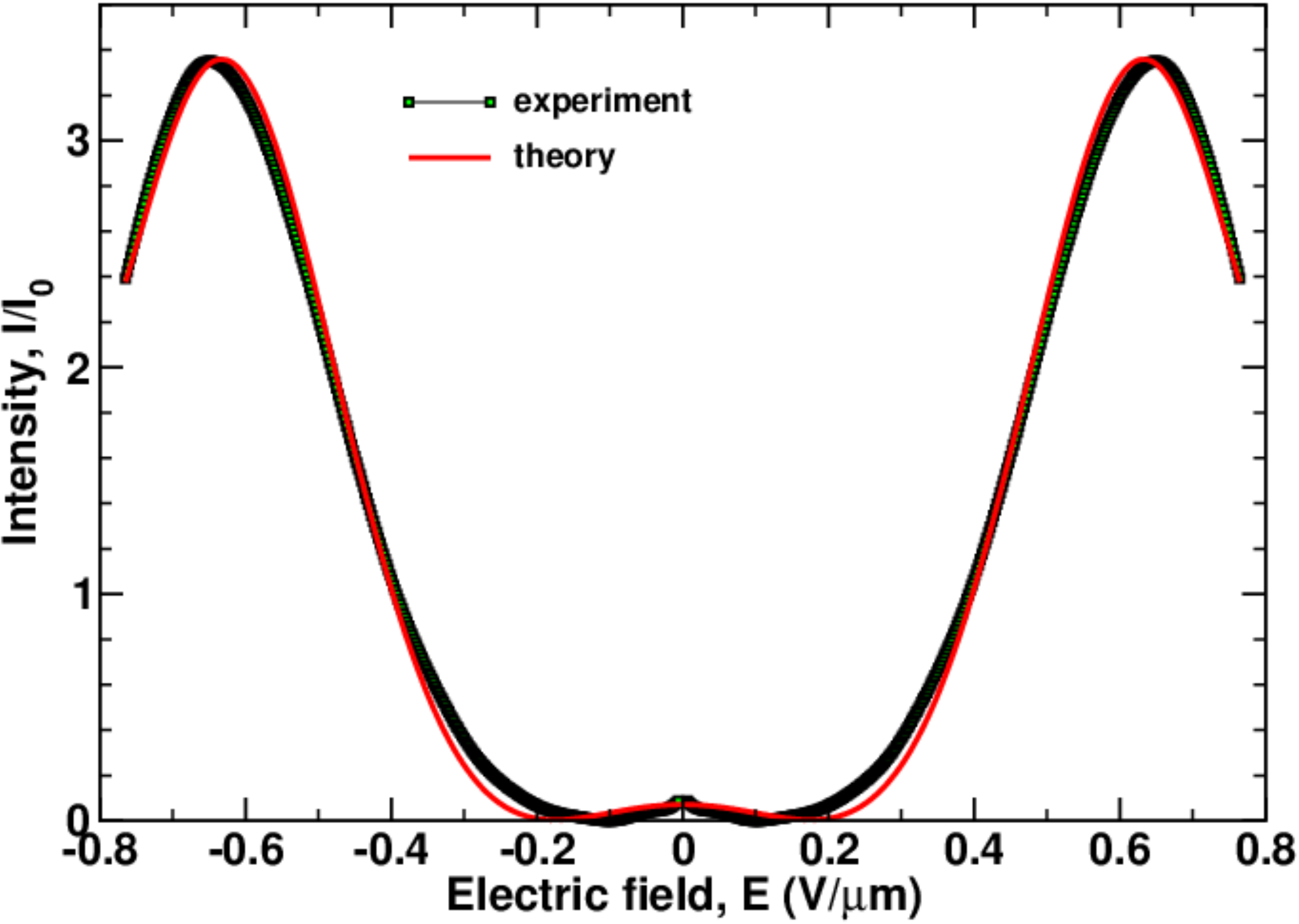}}
\caption{%
Intensity of output wavefield,
$I/I_0$, as a function of the electric field
for the DHFLC cell of thickness
$D\approx 53$~$\mu$m
filled with the FLC mixture FLC-624.
Solid line represents the theoretical curve
computed from~\eqref{eq:intensity}
using the following parameters of the mixture:
$n_{\perp}\approx 1.5$
is the ordinary refractive index,
$n_{\parallel}\approx 1.71$
is the extraordinary refractive index,
$\theta\approx 33^\circ$ is the smectic tilt angle, and 
$r_2\approx 1.03$ is the biaxiality ratio.
}
\label{fig:inten-total}
\end{figure}

Note that the points where the visibility vanishes 
($V=0$) represent the phase
singularities where the Pancharatnam phase is undefined.  
In the
observation plane,  such singularity points
may, under certain conditions, form curves 
that can be visualized as the zero visibility fringes. 
From~\eqref{eq:V-unp}, 
such fringes indicate
the loci of the points 
where the phase retardation
meets the condition of half-wave plates: $\cos(\Delta\Phi/2)=0$. 
So, these singularity
lines might be called the \textit{half-wave fringes}.
Figure~\ref{fig:hw-fringes}
shows the interferogram where 
such fringes arising from variations in cell thickness 
are easily discernible
as zero-contrast stripes separating 
different areas of the interference fringes.
 
\begin{figure}[!tbh]
\centering
\resizebox{80mm}{!}{\includegraphics*{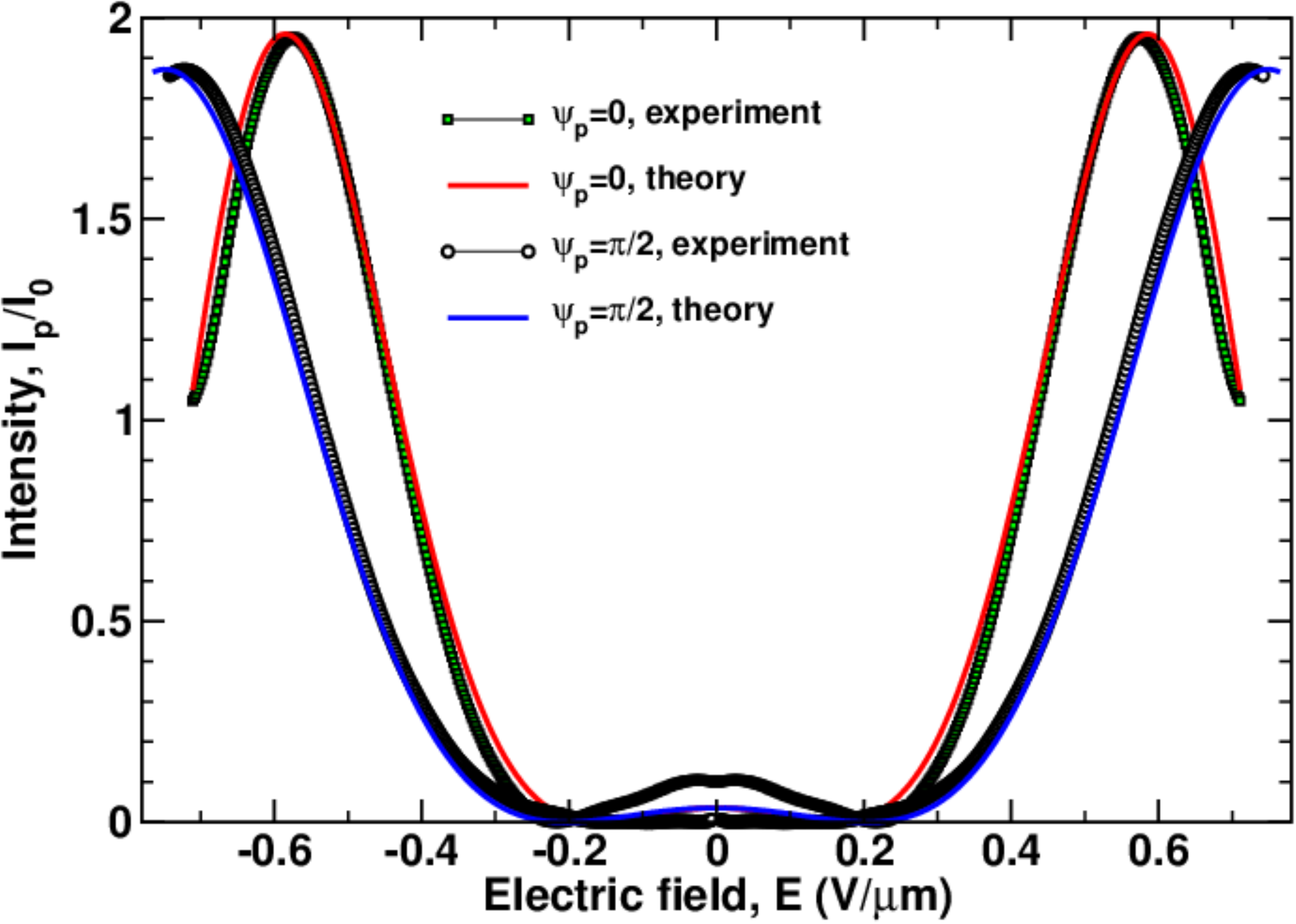}}
\caption{%
Intensity of output wavefield 
$I_p/I_0$, measured as a function of the electric field
after the polarizer with
the transmission axis oriented
along ($\psi_{p}=0$)
and perpendicular ($\psi_p=\pi/2$)
to the helix axis. 
Solid lines represent the theoretical curves
computed from~\eqref{eq:inten-polarizer}
using the parameters listed in the caption of 
Fig.~\ref{fig:inten-total}.
}
\label{fig:inten-polarizer}
\end{figure}

Formulas~(\ref{eq:intensity}) and~(\ref{eq:inten-polarizer})
in combination with the expressions
for $n_{\pm}$ [see Eq.~\eqref{eq:epsilon_pm}] 
and $\psi_\dd$ [see Eq.~\eqref{eq:psi_d}]
can now be used to fit 
the data on electric field dependence of 
the  light intensity measured by the PMT
and
presented in Figs.~\ref{fig:inten-total} and~\ref{fig:inten-polarizer}.
For this purpose, we assume that
the FLC mixture is characterized by the parameters:
$\epsilon_{\perp}\approx 2.22$ ($n_{\perp}\equiv n_o\approx 1.5$), 
$\epsilon_{\parallel}\approx 2.92$ ($n_{\parallel}\equiv n_e\approx 1.71$)
and $\theta=33$\dega.
Then the fitting gives 
the values of ratios: 
$P_s/\chi_E\approx 2.26$~V/\mum\ 
and $r_2=\epsilon_2/\epsilon_{\perp}=1.03$
that are regarded as the fitting parameters.
The theoretical curves are
shown in 
Figs.~\ref{fig:inten-total} and~\ref{fig:inten-polarizer}.
Interestingly, the value of the biaxiality ratio
differs from unity and thus the optical anisotropy of
the mixture appears to be weakly biaxial.
Similar result was reported in our previous study~\cite{Kiselev:pre:2015}.

\section{Discussion and conclusion}
\label{sec:conclusion}

In this paper, 
we have studied 
electric-field-induced modulation of unpolarized light
in the DHFLC  with subwavelength helix pitch 
using the experimental technique 
based on the Mach-Zehnder interferometer.
Such modulation 
occurs under the action of
the voltage applied across the cell
and manifests itself
in the electrically dependent shift and contrast
of the interference pattern. 

The shift and the visibility are 
found to be governed
by the Pancharatnam phase and 
the phase retardation, respectively.
The distinctive features of 
the electro-optic response
in the regime of unpolarized illumination
are:
(a)~insensitivity to
the rotation of in-plane optical axes
(this property was initially regarded 
as a requirement for pure phase modulation);
and
(b)~modulation
at subkiloherz operating frequencies
(similar result was reported in~\cite{Kiselev:ol:2014}
for vertically aligned DHFLCs).
So, modulation of unpolarized light
driven by the orientational Kerr effect
might be of considerable importance
for applications 
in photonic devices
such as deflectors, switchable gratings and wave-front correctors.

Our concluding remark concerns
the Pancharatnam phase
that, for the mixed polarization state~\eqref{eq:rho_0},
is given by formulas~\eqref{eq:PhiP} and~\eqref{eq:F_P}.
The electrically induced shift of the interference fringes
is naturally associated with this phase.
We can also 
apply the approach of Refs.~\cite{Sjoqvist:prl:2000,Tong:prl:2004}
to derive the expression
for the geometric phase $\Phi_g$
~---~the so-called Sj\"oqvist's phase
~--~which is the part of $\Phi_P$
obtained by excluding the dynamical contribution.
Such phase is an important
characteristics which is determined solely
by the geometry of the path in the space of polarization states.

In particular, 
when the transmission
matrix is unitary 
[see Eq.~\eqref{eq:t_pm_approx}],
it is rather straightforward to show that
the geometric phase
can be written in the following form:
\begin{align}
  \label{eq:Phi_g}
&
  \Phi_g=\arg(\Re F_g+i P_0 \Im F_g),
\\
&
\label{eq:F_g}
F_g=[\cos(\Delta\Phi)+i \sin(\Delta\Phi)\cos\psi_0]
\e^{-i\Delta\Phi\cos\psi_0}.
\end{align}
From Eqs.~\eqref{eq:Phi_g} and~\eqref{eq:F_g},
it can be inferred that, similar to the visibility,
$V=|F_P|$, 
the geometric phase $\Phi_g$ depends on 
the phase retardation $2\Delta\Phi$,
and the angle between the unit Stokes vectors,
$\uvc{s}_0$ and $\uvc{s}_d$,
characterizing the input wave and the light linearly
polarized along the optical axis $\uvc{d}_{+}$, respectively: 
$\cos\psi_0=\uvc{s}_0\cdot\uvc{s}_d$.
It comes as no surprise that, for unpolarized incident light, 
the geometric phase~\eqref{eq:Phi_g} vanishes.
Similar remark applies to the case of circular polarized light
where $\cos\psi_0=0$.
A more general case of non-unitary evolution requires
a more sophisticated and comprehensive analysis of 
the electrically dependent geometric phases
generated by DHFLC cells.
Such analysis is beyond the scope of this paper
and its results will be published elsewhere.

\begin{acknowledgments} 
This work is supported by the 
RFBR grants 16-02-00441~A, 16-29-14012~ofi\_m
and 16-42-630773~p\_a.
A.D.K. acknowledges partial financial support
from the Government of the Russian Federation 
(Grant No. 074-U01).
\end{acknowledgments} 

\appendix

\section{Dielectric tensor
of homogenized DHFLC cells}
\label{sec:diel-tensor-dhf}

In this Appendix we recapitulate 
the key results on the effective dielectric
tensor of the DHFLC cells.
More details on these results can 
be found in
Refs.~\cite{Kiselev:pre:2011,Kiselev:pre:2013,Kiselev:pre:2:2014}.

According to Ref.~\cite{Kiselev:pre:2011},
when the pitch-to-wavelength ratio $P/\lambda$
is sufficiently small,
$P/\lambda<1$,
the effective dielectric tensor
\begin{align}
  \label{eq:eff-diel-tensor}
&
\bs{\varepsilon}_{\eff}=
\begin{pmatrix}
  \epsilon_{xx}^{(\eff)} & \epsilon_{xy}^{(\eff)}& \epsilon_{xz}^{(\eff)}\\
\epsilon_{yx}^{(\eff)}& \epsilon_{yy}^{(\eff)} & \epsilon_{yz}^{(\eff)}\\
\epsilon_{zx}^{(\eff)} & \epsilon_{zy}^{(\eff)} & \epsilon_{zz}^{(\eff)}
\end{pmatrix}  
\end{align}
can be expressed in terms of the averages
over the pitch of
the uniform lying distorted FLC helical structure
    \begin{align}
&
   \eta_{zz}=
\avr{\epsilon_{zz}^{-1}}
=
\epsilon_{0}^{-1}
\avr{[1+u_1 d_z^2+u_2 p_z^2]^{-1}},
\label{eq:eta}
\\
&
    \beta_{z\alpha}=
\avr{\epsilon_{z\alpha}/\epsilon_{zz}}
=
\left\langle
\frac{u_1 d_z d_\alpha+u_2 p_z p_\alpha}{1+u_1 d_z^2+u_2 p_z^2}
\right\rangle,
\label{eq:beta}
  \end{align}
where
$\avr{\ldots}\equiv\avr{\ldots}_{\phi}=(2\pi)^{-1}\int_{0}^{2\pi}\ldots\dd\phi$
and $\alpha\in\{x,y\}$,
as follows:
\begin{align}
&
\epsilon_{zz}^{(\eff)}
=1/\eta_{zz},
\quad
\epsilon_{z\alpha}^{(\eff)}
=\beta_{z\alpha}/\eta_{zz},
\notag
\\
&
\epsilon_{\alpha\beta}^{(\eff)}
=\avr{\epsilon_{\alpha\beta}^{(P)}}+
\beta_{z\alpha} \beta_{z\beta}/\eta_{zz},
\label{eq:elements-eff-diel-tensor}
  \end{align}
where
$\avr{\epsilon_{\alpha\beta}^{(P)}}$
are the components of 
the averaged tensor
$\avr{\bs{\varepsilon}_P}$,
\begin{align}
&
\avr{\epsilon_{\alpha\beta}^{(P)}}=
\left\langle
 \epsilon_{\alpha\,\beta}-\frac{ \epsilon_{\alpha\,z}
   \epsilon_{z\,\beta}}{ \epsilon_{zz}}
\right\rangle
\notag
\\
&
=
\epsilon_{0}
\left\langle
\delta_{\alpha\beta}+\frac{u_1 d_\alpha d_\beta+u_2 p_\alpha
  p_\beta+u_1 u_2 q_\alpha q_\beta}{1+u_1  d_z^2+u_2  p_z^2}\,
\right\rangle,
\label{eq:in-plane}
\\
&
q_{\alpha}=p_z d_{\alpha}-d_z p_{\alpha},
\quad 
\alpha,\beta\in\{x,y\},
\label{eq:q-vector}
\end{align}
describing effective in-plane anisotropy
that governs propagation of normally incident plane waves.

The general
formulas~\eqref{eq:eta}-\eqref{eq:q-vector}
give  the zeroth-order approximation
for homogeneous models
describing the optical properties of
short-pitch DHFLCs.
These formulas 
can be used to derive the effective optical tensor of
the homogenized short-pitch DHFLC cell 
for both vertically and planar aligned
FLC helices~\cite{Kiselev:pre:2011,Kiselev:pre:2013,Kiselev:pre:2:2014}.

We concentrate on the geometry
of the uniform lying DHFLC helix
shown in Fig.~\ref{fig:cell}.
For this geometry, 
the effective dielectric tensor can be written
in the following form~\cite{Kiselev:pre:2:2014}:
\begin{align}
  \label{eq:eff-diel-planar}
  &
\bs{\varepsilon}_{\eff}=
\begin{pmatrix}
  \epsilon_h+\gamma_{xx}\alpha_E^2 & \gamma_{xy} \alpha_E& 0\\
  \gamma_{xy} \alpha_E,& \epsilon_p+\gamma_{yy}\alpha_E^2 & 0\\
0 & 0 & \epsilon_p-\gamma_{yy}\alpha_E^2
\end{pmatrix},
\end{align}
where, following Ref.~\cite{Kiselev:pre:2013},
we have 
introduced the electric field parameter
\begin{align}
  \label{eq:alpha_E}
\alpha_E=\chi_E E/P_s  
\end{align}
proportional to
the dielectric susceptibility of 
the Goldstone mode~\cite{Carlsson:pra:1990,Urbanc:ferro:1991}:
$\chi_E=\partial \avr{P_z}/\partial E$ with 
$P_z=P_s\cos\phi$.

The \textit{zero-field dielectric constants}, 
$\epsilon_h$ and $\epsilon_p$,
that enter the tensor~\eqref{eq:eff-diel-planar} are given by
\begin{subequations}
  \label{eq:epsilon_ph}
\begin{align}
&
\label{eq:epsilon_h}
\epsilon_h
/\epsilon_{\perp}
=(n_h/n_{\perp})^2=
r_2^{-1/2}
\biggl\{
\sqrt{r_2}
\notag
\\
&
+u_1 \cos^2\theta
\left(
\frac{r_2-1}{\sqrt{u}+\sqrt{r_2}}+u^{-1/2}
\right)
\biggr\},
\\
&
\label{eq:epsilon_p}
\epsilon_p/\epsilon_{\perp}=
(n_p/n_{\perp})^2=
\sqrt{r_2 u},
\\
&
\label{eq:u}
u=
u_1\sin^2\theta+1.
\end{align}
\end{subequations}
Similar results for 
the \textit{coupling coefficients}
$\gamma_{xx}$,
$\gamma_{yy}$ and $\gamma_{xy}$
read
\begin{subequations}
  \label{eq:coupling-coeffs}
   \begin{align}
&
\label{eq:gxx-u}
\gamma_{xx}/\epsilon_{\perp}=
\frac{3\sqrt{r_2/u}}{(\sqrt{u}+\sqrt{r_2})^2}
(u_1\cos\theta\sin\theta)^2,
\\
&
\label{eq:gyy-u}
\gamma_{yy}/\epsilon_{\perp}=
\frac{3\sqrt{r_2u}}{(\sqrt{u}+\sqrt{r_2})^2}
(u-r_2),
\\
&
\label{eq:gxy-u}
\gamma_{xy}/\epsilon_{\perp}=
\frac{2\sqrt{r_2}}{\sqrt{u}+\sqrt{r_2}}
u_1\cos\theta\sin\theta.
    \end{align}
\end{subequations}

In Ref.~\cite{Kiselev:pre:2:2014}, 
the results~\eqref{eq:eff-diel-planar}--\eqref{eq:gxy-u}
were derived by using
the averaging technique that 
allows high-order corrections to the dielectric tensor 
to be accurately estimated
and
improves agreement between the theory and the experimental
data in the high-field region.

The dielectric tensor~\eqref{eq:eff-diel-planar}
is characterized by the three generally different principal values
(eigenvalues)
and the corresponding \textit{optical axes} (eigenvectors)
as follows
\begin{align}
&
  \label{eq:eff-diel-diag-planar}
  \bs{\varepsilon}_{\eff}=
\epsilon_z \uvc{z}\otimes\uvc{z}
+\epsilon_{+} \uvc{d}_{+}\otimes\uvc{d}_{+}
+\epsilon_{-} \uvc{d}_{-}\otimes\uvc{d}_{-},
\\
&
\label{eq:epsilon_z}
\epsilon_{z}=n_{z}^{\,2}=\epsilon_{zz}^{(\eff)}=\epsilon_p-\gamma_{yy}\alpha_E^2,
\\
&
\label{eq:epsilon_pm}
\epsilon_{\pm}=n_{\pm}^{\,2}=\bar{\epsilon}\pm\sqrt{[\Delta\epsilon]^2+[\gamma_{xy} \alpha_{E}]^2}
\end{align}
where
\begin{align}
&
  \label{eq:epsilon_avr}
  \bar{\epsilon}=(\epsilon_{xx}^{(\eff)}+\epsilon_{yy}^{(\eff)})/2
=\bar{\epsilon}_0+(\gamma_{xx}+\gamma_{yy})\alpha_{E}^2/2,
\\
&
  \label{eq:Delta_epsilon}
 \Delta\epsilon=(\epsilon_{xx}^{(\eff)}-\epsilon_{yy}^{(\eff)})/2=
\Delta\epsilon_0+(\gamma_{xx}-\gamma_{yy})\alpha_{E}^2/2,
\\
&
  \label{eq:Delta_epsilon0}
\bar{\epsilon}_0=(\epsilon_{h}+\epsilon_{p})/2,\:
\Delta\epsilon_0=(\epsilon_{h}-\epsilon_{p})/2.
\end{align}
The in-plane optical axes are given by
\begin{align}
&
  \label{eq:d_plus}
  \uvc{d}_{+}=\cos\psi_d\,\uvc{x}+
\sin\psi_d\,\uvc{y},
\quad
  \uvc{d}_{-}=\uvc{z}\times  \uvc{d}_{+},
\\
&
  \label{eq:psi_d}
2\psi_\dd =\arg[\Delta\epsilon +i \gamma_{xy} \alpha_E].
\end{align}

From Eq.~\eqref{eq:eff-diel-planar}, it is clear that
the zero-field dielectric tensor is
uniaxially anisotropic with the optical axis directed along the
twisting axis $\uvc{h}=\uvc{x}$. 
The applied electric field
changes the principal values 
(see Eqs.~\eqref{eq:epsilon_z} and~\eqref{eq:epsilon_pm})
so that 
the electric-field-induced anisotropy is generally biaxial.
In addition, the in-plane principal optical axes are rotated about the
vector of electric field, $\vc{E}\parallel \uvc{z}$,
by the angle $\psi_\dd$ given in Eq.~\eqref{eq:psi_d}.


\begin{thebibliography}{23}%
\makeatletter
\providecommand \@ifxundefined [1]{%
 \@ifx{#1\undefined}
}%
\providecommand \@ifnum [1]{%
 \ifnum #1\expandafter \@firstoftwo
 \else \expandafter \@secondoftwo
 \fi
}%
\providecommand \@ifx [1]{%
 \ifx #1\expandafter \@firstoftwo
 \else \expandafter \@secondoftwo
 \fi
}%
\providecommand \natexlab [1]{#1}%
\providecommand \enquote  [1]{``#1''}%
\providecommand \bibnamefont  [1]{#1}%
\providecommand \bibfnamefont [1]{#1}%
\providecommand \citenamefont [1]{#1}%
\providecommand \href@noop [0]{\@secondoftwo}%
\providecommand \href [0]{\begingroup \@sanitize@url \@href}%
\providecommand \@href[1]{\@@startlink{#1}\@@href}%
\providecommand \@@href[1]{\endgroup#1\@@endlink}%
\providecommand \@sanitize@url [0]{\catcode `\\12\catcode `\$12\catcode
  `\&12\catcode `\#12\catcode `\^12\catcode `\_12\catcode `\%12\relax}%
\providecommand \@@startlink[1]{}%
\providecommand \@@endlink[0]{}%
\providecommand \url  [0]{\begingroup\@sanitize@url \@url }%
\providecommand \@url [1]{\endgroup\@href {#1}{\urlprefix }}%
\providecommand \urlprefix  [0]{URL }%
\providecommand \Eprint [0]{\href }%
\providecommand \doibase [0]{http://dx.doi.org/}%
\providecommand \selectlanguage [0]{\@gobble}%
\providecommand \bibinfo  [0]{\@secondoftwo}%
\providecommand \bibfield  [0]{\@secondoftwo}%
\providecommand \translation [1]{[#1]}%
\providecommand \BibitemOpen [0]{}%
\providecommand \bibitemStop [0]{}%
\providecommand \bibitemNoStop [0]{.\EOS\space}%
\providecommand \EOS [0]{\spacefactor3000\relax}%
\providecommand \BibitemShut  [1]{\csname bibitem#1\endcsname}%
\let\auto@bib@innerbib\@empty
\bibitem [{\citenamefont {Efron}(1995)}]{Efron:bk:1995}%
  \BibitemOpen
  \bibinfo {editor} {\bibfnamefont {Uzi}\ \bibnamefont {Efron}},\ ed.,\
  \href@noop {} {{\selectlanguage {english}\emph {\bibinfo {title} {Spatial
  Light Modulator Technology: Materials, Applications, and Devices}}}}\
  (\bibinfo  {publisher} {Marcell Dekker},\ \bibinfo {address} {NY},\ \bibinfo
  {year} {1995})\ p.\ \bibinfo {pages} {665}\BibitemShut {NoStop}%
\bibitem [{\citenamefont {Oswald}\ and\ \citenamefont
  {Pieranski}(2006)}]{Oswald:bk:2006}%
  \BibitemOpen
  \bibfield  {author} {\bibinfo {author} {\bibfnamefont {Patrick}\ \bibnamefont
  {Oswald}}\ and\ \bibinfo {author} {\bibfnamefont {Pawel}\ \bibnamefont
  {Pieranski}},\ }\href@noop {} {{\selectlanguage {english}\emph {\bibinfo
  {title} {Smectic and Columnar Liquid Crystals: Concepts and Physical
  Properies Illustrated by Experiments}}}},\ The Liquid Crystals Book Series\
  (\bibinfo  {publisher} {Taylor \& Francis Group},\ \bibinfo {address}
  {London},\ \bibinfo {year} {2006})\ p.\ \bibinfo {pages} {690}\BibitemShut
  {NoStop}%
\bibitem [{\citenamefont {Love}\ and\ \citenamefont
  {Bhandari}(1994)}]{Love:optcom:1994}%
  \BibitemOpen
  \bibfield  {author} {\bibinfo {author} {\bibfnamefont {G.~D.}\ \bibnamefont
  {Love}}\ and\ \bibinfo {author} {\bibfnamefont {R.}~\bibnamefont
  {Bhandari}},\ }\bibfield  {title} {{\selectlanguage {english}\enquote
  {\bibinfo {title} {Optical properties of {QHQ} ferroelectric liquid crystal
  phase modulator},}\ }}\href@noop {} {\bibfield  {journal} {\bibinfo
  {journal} {Opt. Commun.}\ }\textbf {\bibinfo {volume} {110}},\ \bibinfo
  {pages} {475--478} (\bibinfo {year} {1994})}\BibitemShut {NoStop}%
\bibitem [{\citenamefont {Barnik}\ \emph {et~al.}(1987)\citenamefont {Barnik},
  \citenamefont {Baikalov}, \citenamefont {Chigrinov},\ and\ \citenamefont
  {Pozhidaev}}]{Barnik:mclc:1987}%
  \BibitemOpen
  \bibfield  {author} {\bibinfo {author} {\bibfnamefont {M.~I.}\ \bibnamefont
  {Barnik}}, \bibinfo {author} {\bibfnamefont {V.~A.}\ \bibnamefont
  {Baikalov}}, \bibinfo {author} {\bibfnamefont {V.~G.}\ \bibnamefont
  {Chigrinov}}, \ and\ \bibinfo {author} {\bibfnamefont {E.~P.}\ \bibnamefont
  {Pozhidaev}},\ }\bibfield  {title} {{\selectlanguage {english}\enquote
  {\bibinfo {title} {Electrooptics of a thin ferroeiectric smectic {$C^*$}
  liquid crystal layer},}\ }}\href@noop {} {\bibfield  {journal} {\bibinfo
  {journal} {Mol. Cryst. Liq. Cryst.}\ }\textbf {\bibinfo {volume} {143}},\
  \bibinfo {pages} {101--112} (\bibinfo {year} {1987})}\BibitemShut {NoStop}%
\bibitem [{\citenamefont {Pozhidaev}\ \emph {et~al.}(1988)\citenamefont
  {Pozhidaev}, \citenamefont {Osipov}, \citenamefont {Chigrinov}, \citenamefont
  {Baikalov}, \citenamefont {Blinov},\ and\ \citenamefont
  {Beresnev}}]{Pozhidaev:jetp:1988}%
  \BibitemOpen
  \bibfield  {author} {\bibinfo {author} {\bibfnamefont {E.~P.}\ \bibnamefont
  {Pozhidaev}}, \bibinfo {author} {\bibfnamefont {M.~A.}\ \bibnamefont
  {Osipov}}, \bibinfo {author} {\bibfnamefont {V.~G.}\ \bibnamefont
  {Chigrinov}}, \bibinfo {author} {\bibfnamefont {V.~A.}\ \bibnamefont
  {Baikalov}}, \bibinfo {author} {\bibfnamefont {L.~M.}\ \bibnamefont
  {Blinov}}, \ and\ \bibinfo {author} {\bibfnamefont {L.~A.}\ \bibnamefont
  {Beresnev}},\ }\bibfield  {title} {{\selectlanguage {english}\enquote
  {\bibinfo {title} {Rotational viscosity of the smectic {$C^\star$} phase of
  ferroelectric liquid crystals},}\ }}\href@noop {} {\bibfield  {journal}
  {\bibinfo  {journal} {Zh. Eksp. Teor. Fiz.}\ }\textbf {\bibinfo {volume}
  {94}},\ \bibinfo {pages} {125--132} (\bibinfo {year} {1988})}\BibitemShut
  {NoStop}%
\bibitem [{\citenamefont {Pozhidaev}\ \emph {et~al.}(2013)\citenamefont
  {Pozhidaev}, \citenamefont {Kiselev}, \citenamefont {Srivastava},
  \citenamefont {Chigrinov}, \citenamefont {Kwok},\ and\ \citenamefont
  {Minchenko}}]{Kiselev:pre:2013}%
  \BibitemOpen
  \bibfield  {author} {\bibinfo {author} {\bibfnamefont {Evgeny~P.}\
  \bibnamefont {Pozhidaev}}, \bibinfo {author} {\bibfnamefont {Alexei~D.}\
  \bibnamefont {Kiselev}}, \bibinfo {author} {\bibfnamefont {Abhishek~Kumar}\
  \bibnamefont {Srivastava}}, \bibinfo {author} {\bibfnamefont {Vladimir~G.}\
  \bibnamefont {Chigrinov}}, \bibinfo {author} {\bibfnamefont {Hoi-Sing}\
  \bibnamefont {Kwok}}, \ and\ \bibinfo {author} {\bibfnamefont {Maxim~V.}\
  \bibnamefont {Minchenko}},\ }\bibfield  {title} {{\selectlanguage
  {english}\enquote {\bibinfo {title} {Orientational {K}err effect and phase
  modulation of light in deformed-helix ferroelectric liquid crystals with
  subwavelength pitch},}\ }}\href@noop {} {\bibfield  {journal} {\bibinfo
  {journal} {Phys. Rev. E}\ }\textbf {\bibinfo {volume} {87}},\ \bibinfo
  {pages} {052502} (\bibinfo {year} {2013})}\BibitemShut {NoStop}%
\bibitem [{\citenamefont {Pozhidaev}\ \emph {et~al.}(2014)\citenamefont
  {Pozhidaev}, \citenamefont {Srivastava}, \citenamefont {Kiselev},
  \citenamefont {Chigrinov}, \citenamefont {Vashchenko}, \citenamefont
  {Krivoshey}, \citenamefont {Minchenko},\ and\ \citenamefont
  {Kwok}}]{Kiselev:ol:2014}%
  \BibitemOpen
  \bibfield  {author} {\bibinfo {author} {\bibfnamefont {Evgeny~P.}\
  \bibnamefont {Pozhidaev}}, \bibinfo {author} {\bibfnamefont {Abhishek~Kumar}\
  \bibnamefont {Srivastava}}, \bibinfo {author} {\bibfnamefont {Alexei~D.}\
  \bibnamefont {Kiselev}}, \bibinfo {author} {\bibfnamefont {Vladimir~G.}\
  \bibnamefont {Chigrinov}}, \bibinfo {author} {\bibfnamefont {Valery~V.}\
  \bibnamefont {Vashchenko}}, \bibinfo {author} {\bibfnamefont {Alexander~V.}\
  \bibnamefont {Krivoshey}}, \bibinfo {author} {\bibfnamefont {Maxim~V.}\
  \bibnamefont {Minchenko}}, \ and\ \bibinfo {author} {\bibfnamefont
  {Hoi-Sing}\ \bibnamefont {Kwok}},\ }\bibfield  {title} {{\selectlanguage
  {english}\enquote {\bibinfo {title} {Enhanced orientational {K}err effect in
  vertically aligned deformed helix ferroelectric liquid crystals},}\
  }}\href@noop {} {\bibfield  {journal} {\bibinfo  {journal} {Optics Letters}\
  }\textbf {\bibinfo {volume} {39}},\ \bibinfo {pages} {2900--2903} (\bibinfo
  {year} {2014})}\BibitemShut {NoStop}%
\bibitem [{\citenamefont {Blinov}\ \emph {et~al.}(2005)\citenamefont {Blinov},
  \citenamefont {Palto}, \citenamefont {Pozhidaev}, \citenamefont {Bobylev},
  \citenamefont {Shoshin}, \citenamefont {Andreev}, \citenamefont {Podgornov},\
  and\ \citenamefont {Haase}}]{Blinov:pre:2005}%
  \BibitemOpen
  \bibfield  {author} {\bibinfo {author} {\bibfnamefont {L.~M.}\ \bibnamefont
  {Blinov}}, \bibinfo {author} {\bibfnamefont {S.~P.}\ \bibnamefont {Palto}},
  \bibinfo {author} {\bibfnamefont {E.~P.}\ \bibnamefont {Pozhidaev}}, \bibinfo
  {author} {\bibfnamefont {Yu.~P.}\ \bibnamefont {Bobylev}}, \bibinfo {author}
  {\bibfnamefont {V.~M.}\ \bibnamefont {Shoshin}}, \bibinfo {author}
  {\bibfnamefont {A.~L.}\ \bibnamefont {Andreev}}, \bibinfo {author}
  {\bibfnamefont {F.~V.}\ \bibnamefont {Podgornov}}, \ and\ \bibinfo {author}
  {\bibfnamefont {W.}~\bibnamefont {Haase}},\ }\bibfield  {title}
  {{\selectlanguage {english}\enquote {\bibinfo {title} {High frequency
  hysteresis-free switching in thin layers smectic-{$C^*$} ferroelectric liquid
  crystals},}\ }}\href@noop {} {\bibfield  {journal} {\bibinfo  {journal}
  {Phys. Rev. E}\ }\textbf {\bibinfo {volume} {71}},\ \bibinfo {pages} {071715}
  (\bibinfo {year} {2005})}\BibitemShut {NoStop}%
\bibitem [{\citenamefont {Pozhidaev}\ \emph {et~al.}(2012)\citenamefont
  {Pozhidaev}, \citenamefont {Chigrinov}, \citenamefont {Murauski},
  \citenamefont {Molkin}, \citenamefont {Tao},\ and\ \citenamefont
  {Kwok}}]{Pozhidaev:jsid:2012}%
  \BibitemOpen
  \bibfield  {author} {\bibinfo {author} {\bibfnamefont {Eugene}\ \bibnamefont
  {Pozhidaev}}, \bibinfo {author} {\bibfnamefont {Vladimir}\ \bibnamefont
  {Chigrinov}}, \bibinfo {author} {\bibfnamefont {Anatoli}\ \bibnamefont
  {Murauski}}, \bibinfo {author} {\bibfnamefont {Vadim}\ \bibnamefont
  {Molkin}}, \bibinfo {author} {\bibfnamefont {Du}~\bibnamefont {Tao}}, \ and\
  \bibinfo {author} {\bibfnamefont {Hoi-Sing}\ \bibnamefont {Kwok}},\
  }\bibfield  {title} {{\selectlanguage {english}\enquote {\bibinfo {title}
  {V-shaped electro-optical mode based on deformed-helix ferroelectric liquid
  crystal with subwavelength pitch},}\ }}\href@noop {} {\bibfield  {journal}
  {\bibinfo  {journal} {Journal of the SID}\ }\textbf {\bibinfo {volume}
  {20}},\ \bibinfo {pages} {273--278} (\bibinfo {year} {2012})}\BibitemShut
  {NoStop}%
\bibitem [{\citenamefont {Kotova}\ \emph {et~al.}(2015)\citenamefont {Kotova},
  \citenamefont {Samagin}, \citenamefont {Pozhidaev},\ and\ \citenamefont
  {Kiselev}}]{Kiselev:pre:2015}%
  \BibitemOpen
  \bibfield  {author} {\bibinfo {author} {\bibfnamefont {Svetlana~P.}\
  \bibnamefont {Kotova}}, \bibinfo {author} {\bibfnamefont {Sergey~A.}\
  \bibnamefont {Samagin}}, \bibinfo {author} {\bibfnamefont {Evgeny~P.}\
  \bibnamefont {Pozhidaev}}, \ and\ \bibinfo {author} {\bibfnamefont
  {Alexei~D.}\ \bibnamefont {Kiselev}},\ }\bibfield  {title} {{\selectlanguage
  {english}\enquote {\bibinfo {title} {Light modulation in planar aligned
  short-pitch deformed-helix ferroelectric liquid crystals},}\ }}\href@noop {}
  {\bibfield  {journal} {\bibinfo  {journal} {Phys. Rev. E}\ }\textbf {\bibinfo
  {volume} {92}},\ \bibinfo {pages} {062502} (\bibinfo {year}
  {2015})}\BibitemShut {NoStop}%
\bibitem [{\citenamefont {Kiselev}\ \emph {et~al.}(2011)\citenamefont
  {Kiselev}, \citenamefont {Pozhidaev}, \citenamefont {Chigrinov},\ and\
  \citenamefont {Kwok}}]{Kiselev:pre:2011}%
  \BibitemOpen
  \bibfield  {author} {\bibinfo {author} {\bibfnamefont {Alexei~D.}\
  \bibnamefont {Kiselev}}, \bibinfo {author} {\bibfnamefont {Eugene~P.}\
  \bibnamefont {Pozhidaev}}, \bibinfo {author} {\bibfnamefont {Vladimir~G.}\
  \bibnamefont {Chigrinov}}, \ and\ \bibinfo {author} {\bibfnamefont
  {Hoi-Sing}\ \bibnamefont {Kwok}},\ }\bibfield  {title} {{\selectlanguage
  {english}\enquote {\bibinfo {title} {Polarization-gratings approach to
  deformed-helix ferroelectric liquid crystals with subwavelength pitch},}\
  }}\href@noop {} {\bibfield  {journal} {\bibinfo  {journal} {Phys. Rev. E}\
  }\textbf {\bibinfo {volume} {83}},\ \bibinfo {pages} {031703} (\bibinfo
  {year} {2011})}\BibitemShut {NoStop}%
\bibitem [{\citenamefont {Kiselev}\ and\ \citenamefont
  {Chigrinov}(2014)}]{Kiselev:pre:2:2014}%
  \BibitemOpen
  \bibfield  {author} {\bibinfo {author} {\bibfnamefont {Alexei~D.}\
  \bibnamefont {Kiselev}}\ and\ \bibinfo {author} {\bibfnamefont {Vladimir~G.}\
  \bibnamefont {Chigrinov}},\ }\bibfield  {title} {{\selectlanguage
  {english}\enquote {\bibinfo {title} {Optics of short-pitch deformed-helix
  ferroelectric liquid crystals: Symmetries, exceptional points, and
  polarization-resolved angular patterns},}\ }}\href {\doibase
  10.1103/PhysRevE.90.042504} {\bibfield  {journal} {\bibinfo  {journal} {Phys.
  Rev. E}\ }\textbf {\bibinfo {volume} {90}},\ \bibinfo {pages} {042504}
  (\bibinfo {year} {2014})}\BibitemShut {NoStop}%
\bibitem [{\citenamefont {Kiselev}\ \emph {et~al.}(2008)\citenamefont
  {Kiselev}, \citenamefont {Vovk}, \citenamefont {Egorov},\ and\ \citenamefont
  {Chigrinov}}]{Kiselev:pra:2008}%
  \BibitemOpen
  \bibfield  {author} {\bibinfo {author} {\bibfnamefont {A.~D.}\ \bibnamefont
  {Kiselev}}, \bibinfo {author} {\bibfnamefont {R.~G.}\ \bibnamefont {Vovk}},
  \bibinfo {author} {\bibfnamefont {R.~I.}\ \bibnamefont {Egorov}}, \ and\
  \bibinfo {author} {\bibfnamefont {V.~G.}\ \bibnamefont {Chigrinov}},\
  }\bibfield  {title} {{\selectlanguage {english}\enquote {\bibinfo {title}
  {Polarization-resolved angular patterns of nematic liquid crystal cells:
  {T}opological events driven by incident light polarization},}\ }}\href@noop
  {} {\bibfield  {journal} {\bibinfo  {journal} {Phys. Rev. A}\ }\textbf
  {\bibinfo {volume} {78}},\ \bibinfo {pages} {033815} (\bibinfo {year}
  {2008})}\BibitemShut {NoStop}%
\bibitem [{\citenamefont {Mandel}\ and\ \citenamefont
  {Wolf}(1995)}]{Mandl:bk:1995}%
  \BibitemOpen
  \bibfield  {author} {\bibinfo {author} {\bibfnamefont {Leonard}\ \bibnamefont
  {Mandel}}\ and\ \bibinfo {author} {\bibfnamefont {Emil}\ \bibnamefont
  {Wolf}},\ }\href@noop {} {{\selectlanguage {english}\emph {\bibinfo {title}
  {Optical Coherence and Quantum Optics}}}}\ (\bibinfo  {publisher} {Cambridge
  University Press},\ \bibinfo {address} {Cambridge},\ \bibinfo {year} {1995})\
  p.\ \bibinfo {pages} {1194}\BibitemShut {NoStop}%
\bibitem [{\citenamefont {Brosseau}(1998)}]{Brosseau:bk:1998}%
  \BibitemOpen
  \bibfield  {author} {\bibinfo {author} {\bibfnamefont {Christian}\
  \bibnamefont {Brosseau}},\ }\href@noop {} {{\selectlanguage {english}\emph
  {\bibinfo {title} {Fundamentals of Polarized Light: A Statistical Optics
  Approach}}}}\ (\bibinfo  {publisher} {Wiley},\ \bibinfo {address} {New
  York},\ \bibinfo {year} {1998})\ p.\ \bibinfo {pages} {424}\BibitemShut
  {NoStop}%
\bibitem [{\citenamefont {Pancharatnam}(1956)}]{Pancharatnam:proc:1956}%
  \BibitemOpen
  \bibfield  {author} {\bibinfo {author} {\bibfnamefont {S.}~\bibnamefont
  {Pancharatnam}},\ }\bibfield  {title} {{\selectlanguage {english}\enquote
  {\bibinfo {title} {Generalized theory of interference and its
  apllications},}\ }}\href@noop {} {\bibfield  {journal} {\bibinfo  {journal}
  {The Proceedings of the Indian Academy of Sciences}\ }\textbf {\bibinfo
  {volume} {Sect. A 44}},\ \bibinfo {pages} {247} (\bibinfo {year}
  {1956})}\BibitemShut {NoStop}%
\bibitem [{\citenamefont {Shapere}\ and\ \citenamefont
  {Wilczek}(1989)}]{Shapere:bk:1989}%
  \BibitemOpen
  \bibfield  {author} {\bibinfo {author} {\bibfnamefont {Alfred}\ \bibnamefont
  {Shapere}}\ and\ \bibinfo {author} {\bibfnamefont {Frank}\ \bibnamefont
  {Wilczek}},\ }\href@noop {} {{\selectlanguage {english}\emph {\bibinfo
  {title} {Geometric Phases in Physics}}}},\ \bibinfo {series} {Advanced Series
  in Mathematical Physics}, Vol.~\bibinfo {volume} {5}\ (\bibinfo  {publisher}
  {World Scientific},\ \bibinfo {address} {Singapore},\ \bibinfo {year}
  {1989})\ p.\ \bibinfo {pages} {509}\BibitemShut {NoStop}%
\bibitem [{\citenamefont {Hariharan}(2005)}]{Hariharan:progr_opt:2005}%
  \BibitemOpen
  \bibfield  {author} {\bibinfo {author} {\bibfnamefont {P.}~\bibnamefont
  {Hariharan}},\ }\bibfield  {title} {{\selectlanguage {english}\enquote
  {\bibinfo {title} {The geometric phase},}\ }}\href@noop {} {\bibfield
  {journal} {\bibinfo  {journal} {Progress in Optics}\ }\textbf {\bibinfo
  {volume} {48}},\ \bibinfo {pages} {293--363} (\bibinfo {year}
  {2005})}\BibitemShut {NoStop}%
\bibitem [{\citenamefont {Barberena}\ \emph {et~al.}(2016)\citenamefont
  {Barberena}, \citenamefont {Ort\'{\i}z}, \citenamefont {Yugra}, \citenamefont
  {Caballero},\ and\ \citenamefont {De~Zela}}]{Barberena:pra:2016}%
  \BibitemOpen
  \bibfield  {author} {\bibinfo {author} {\bibfnamefont {D.}~\bibnamefont
  {Barberena}}, \bibinfo {author} {\bibfnamefont {O.}~\bibnamefont
  {Ort\'{\i}z}}, \bibinfo {author} {\bibfnamefont {Y.}~\bibnamefont {Yugra}},
  \bibinfo {author} {\bibfnamefont {R.}~\bibnamefont {Caballero}}, \ and\
  \bibinfo {author} {\bibfnamefont {F.}~\bibnamefont {De~Zela}},\ }\bibfield
  {title} {{\selectlanguage {english}\enquote {\bibinfo {title} {All-optical
  polarimetric generation of mixed-state single-photon geometric phases},}\
  }}\href {\doibase 10.1103/PhysRevA.93.013805} {\bibfield  {journal} {\bibinfo
   {journal} {Phys. Rev. A}\ }\textbf {\bibinfo {volume} {93}},\ \bibinfo
  {pages} {013805} (\bibinfo {year} {2016})}\BibitemShut {NoStop}%
\bibitem [{\citenamefont {Sj\"oqvist}\ \emph {et~al.}(2000)\citenamefont
  {Sj\"oqvist}, \citenamefont {Pati}, \citenamefont {Ekert}, \citenamefont
  {Anandan}, \citenamefont {Ericsson}, \citenamefont {Oi},\ and\ \citenamefont
  {Vedral}}]{Sjoqvist:prl:2000}%
  \BibitemOpen
  \bibfield  {author} {\bibinfo {author} {\bibfnamefont {Erik}\ \bibnamefont
  {Sj\"oqvist}}, \bibinfo {author} {\bibfnamefont {Arun~K.}\ \bibnamefont
  {Pati}}, \bibinfo {author} {\bibfnamefont {Artur}\ \bibnamefont {Ekert}},
  \bibinfo {author} {\bibfnamefont {Jeeva~S.}\ \bibnamefont {Anandan}},
  \bibinfo {author} {\bibfnamefont {Marie}\ \bibnamefont {Ericsson}}, \bibinfo
  {author} {\bibfnamefont {Daniel K.~L.}\ \bibnamefont {Oi}}, \ and\ \bibinfo
  {author} {\bibfnamefont {Vlatko}\ \bibnamefont {Vedral}},\ }\bibfield
  {title} {{\selectlanguage {english}\enquote {\bibinfo {title} {Geometric
  phases for mixed states in interferometry},}\ }}\href {\doibase
  10.1103/PhysRevLett.85.2845} {\bibfield  {journal} {\bibinfo  {journal}
  {Phys. Rev. Lett.}\ }\textbf {\bibinfo {volume} {85}},\ \bibinfo {pages}
  {2845--2849} (\bibinfo {year} {2000})}\BibitemShut {NoStop}%
\bibitem [{\citenamefont {Tong}\ \emph {et~al.}(2004)\citenamefont {Tong},
  \citenamefont {Sj\"oqvist}, \citenamefont {Kwek},\ and\ \citenamefont
  {Oh}}]{Tong:prl:2004}%
  \BibitemOpen
  \bibfield  {author} {\bibinfo {author} {\bibfnamefont {D.~M.}\ \bibnamefont
  {Tong}}, \bibinfo {author} {\bibfnamefont {E.}~\bibnamefont {Sj\"oqvist}},
  \bibinfo {author} {\bibfnamefont {L.~C.}\ \bibnamefont {Kwek}}, \ and\
  \bibinfo {author} {\bibfnamefont {C.~H.}\ \bibnamefont {Oh}},\ }\bibfield
  {title} {{\selectlanguage {english}\enquote {\bibinfo {title} {Kinematic
  approach to the mixed state geometric phase in nonunitary evolution},}\
  }}\href {\doibase 10.1103/PhysRevLett.93.080405} {\bibfield  {journal}
  {\bibinfo  {journal} {Phys. Rev. Lett.}\ }\textbf {\bibinfo {volume} {93}},\
  \bibinfo {pages} {080405} (\bibinfo {year} {2004})}\BibitemShut {NoStop}%
\bibitem [{\citenamefont {Carlsson}\ \emph {et~al.}(1990)\citenamefont
  {Carlsson}, \citenamefont {\v{Z}ek\v{s}}, \citenamefont {Filipi\v{c}},\ and\
  \citenamefont {Levstik}}]{Carlsson:pra:1990}%
  \BibitemOpen
  \bibfield  {author} {\bibinfo {author} {\bibfnamefont {T.}~\bibnamefont
  {Carlsson}}, \bibinfo {author} {\bibfnamefont {B.}~\bibnamefont
  {\v{Z}ek\v{s}}}, \bibinfo {author} {\bibfnamefont {C.}~\bibnamefont
  {Filipi\v{c}}}, \ and\ \bibinfo {author} {\bibfnamefont {A.}~\bibnamefont
  {Levstik}},\ }\bibfield  {title} {{\selectlanguage {english}\enquote
  {\bibinfo {title} {Theoretical model of the frequency and temperature
  dependence of the complex dielectric constant of ferroelectric liquid
  crystals near the smectic-{$C^*$}~---~smectic-{$A$} phase transition},}\
  }}\href@noop {} {\bibfield  {journal} {\bibinfo  {journal} {Phys. Rev. A}\
  }\textbf {\bibinfo {volume} {42}},\ \bibinfo {pages} {877--889} (\bibinfo
  {year} {1990})}\BibitemShut {NoStop}%
\bibitem [{\citenamefont {Urbanc}\ \emph {et~al.}(1991)\citenamefont {Urbanc},
  \citenamefont {\v{Z}ek\v{s}},\ and\ \citenamefont
  {Carlsson}}]{Urbanc:ferro:1991}%
  \BibitemOpen
  \bibfield  {author} {\bibinfo {author} {\bibfnamefont {B.}~\bibnamefont
  {Urbanc}}, \bibinfo {author} {\bibfnamefont {B.}~\bibnamefont
  {\v{Z}ek\v{s}}}, \ and\ \bibinfo {author} {\bibfnamefont {T.}~\bibnamefont
  {Carlsson}},\ }\bibfield  {title} {{\selectlanguage {english}\enquote
  {\bibinfo {title} {Nonlinear effects in the dielectric response of
  ferroelectric liquid crystals},}\ }}\href@noop {} {\bibfield  {journal}
  {\bibinfo  {journal} {Ferroelectrics}\ }\textbf {\bibinfo {volume} {113}},\
  \bibinfo {pages} {219--230} (\bibinfo {year} {1991})}\BibitemShut {NoStop}%
\end{thebibliography}

%

\end{document}